\documentclass{aastex} \usepackage{emulateapj5}
\begin{document}

\title{Lupus-TR-3b: A Low-Mass Transiting Hot Jupiter in the Galactic Plane?} \shorttitle{Lupus-TR-3b} \shortauthors{Weldrake et al}

\author{David T F Weldrake\altaffilmark{1}, Daniel D R Bayliss\altaffilmark{2}, Penny D Sackett\altaffilmark{2}, Brandon W Tingley\altaffilmark{3}, Micha\"el Gillon\altaffilmark{4,5} and Johny Setiawan\altaffilmark{1}}

\altaffiltext{1}{Max Planck Institut f\"ur Astronomie, K\"onigstuhl 17, D-69117, Heidelberg, Germany. Email: weldrake@mpia.de}
\altaffiltext{2}{Research School of Astronomy and Astrophysics, Mount Stromlo Observatory, Cotter Road, Weston Creek, ACT 2611, Australia} 
\altaffiltext{3}{Institute d'Astronomy et Astrophysique, Universit\'e Libr\'e de Bruxelles, Belgium} \altaffiltext{4}{Geneva Observatory, 51 Chemin des Maillettes, 1290 Versoix, Switzerland}
\altaffiltext{5}{Institut d'Astrophysique et de G\'eophysique, Universit\'e De Li\'ege,  4000 Li\'ege, Belgium}

\begin{abstract}
We present a strong case for a transiting Hot Jupiter planet identified during a single-field transit survey towards the Lupus Galactic plane. The object, Lupus-TR-3b, transits a V=17.4  K1V host star every 3.91405d.  Spectroscopy and stellar colors indicate a host star with effective temperature 5000 $\pm$ 150K, with a stellar mass and radius of 0.87 $\pm$ 0.04M$_{\odot}$ and 0.82 $\pm$ 0.05R$_{\odot}$, respectively. Limb-darkened transit fitting yields a companion radius of $0.89 \pm 0.07$ R$_{\rm{J}}$ and an orbital inclination of $88.3 ^{+1.3}_{-0.8}$ deg.   Magellan 6.5m MIKE radial velocity measurements reveal a 2.4$\sigma$ K$=$114 $\pm$ 25m/s sinusoidal variation in phase with the transit ephemeris. The resulting mass is 0.81 $\pm$ 0.18M$_{\rm{J}}$ and density 1.4 $\pm$ 0.4g/cm$^{3}$. Y-band PANIC image deconvolution reveal a V$\geqslant$21 red neighbor 0.4$''$ away which, although highly unlikely, we cannot conclusively rule out as a blended binary with current data. However, blend simulations show that only the most unusual binary system can reproduce our observations. This object is very likely a planet, detected from a highly efficient observational strategy. Lupus-TR-3b constitutes the faintest ground-based detection to date, and one of the lowest mass Hot Jupiters known.
\end{abstract}


\section{Introduction}
Transiting extrasolar planets provide unique opportunities to study the structure of exoplanets. Transit photometry, coupled with radial velocities, allows direct measurement of the exoplanetary radius, total mass and density.  Currently $\sim$30 transiting planets are known\footnote{http://exoplanet.eu/}, the majority identified in wide-field transit surveys and confirmed by radial velocities. Due to their easy detection and transit-related bias, nearly all the currently known eclipsing exoplanets are large ($\sim R_{\rm J}$), the exception being GJ 436b \citep{G2007b}, and in close orbits. Transits identified against bright stars (V$\leqslant$12) permit investigation into planetary atmosphere \citep{VM2004,TIN2007} and infrared emission \citep{C2005,D2005}. Deeper surveys using more numerous fainter stars can detect large numbers of planets for statistical studies of exoplanet populations, and additional system planets through large aperture photometry.

Transit surveys uncover many candidates, most of which are astrophysical false-positives which must undergo strict follow-up procedures \citep{OD2006,OD2007}. Distinguishing hierarchical triples and blended binaries \citep{T2004,Mand2005} requires large time investment and careful analysis with high-resolution, large-aperture spectroscopy and photometry.

We present the detection of Lupus-TR-3b, a strong case for a transiting 0.81 $\pm$ 0.18M$_{\rm{J}}$ Hot Jupiter planet identified from a deep, 0.66deg$^{2}$ transit survey towards Lupus. This result required 12.8 hours of Magellan 6.5m time spread over 5 nights (MIKE spectroscopy shared with another project for increased efficiency, and PANIC imagery including overheads), and 1 hour of ANU 2.3m time (DBS spectroscopy) for a target at V$=$17.4.  We detail the likely nature of this object and the overall system parameters.

\section{The Photometric Survey}
We conducted a 52$'$$\times$52$'$ field transit survey centered at RA$=$15$^{\rm{h}}$30$^{\rm{m}}$36.3$^{\rm{s}}$, DEC$=$$-$42$^{\circ}$53$'$53.0$''$, b$=$11$^\circ$ l$=$331.5$^\circ$ towards Lupus, using the Wide Field Imager (WFI) on the Australian National University (ANU) 1m telescope at Siding Spring Observatory (SSO). Seven of the eight WFI CCDs were operational, yielding an effective 0.66deg$^2$ field. We observed with a single broad-band V+R filter for 26 and 27 nights in June 2005 and June 2006, respectively. Survey motivation is threefold: a transit search in its own right, a control field for the earlier globular cluster transit surveys of \citet{W2005,W2007b}, and a feasibility study for the massive 5.7 deg$^2$ SkyMapper transiting planet survey \citep{BS2007}.

The stellar light curves consist of 1783 five-minute exposures in V+R with a mean seeing of 2.2$''$. A total of 110,372 light curves were produced via differential imaging techniques \citep{AL98,Woz2000}, of which 16,134 have suitable photometric quality (total light curve uncertainty $\le$0.025 mag) for the main transit search. Correlated, systematic errors were removed from the light curves using the SYSREM algorithm \citep{T2005}.

From a tandem application of both the BLS \citep{K2002} and the \citet{WS2005} transit detection algorithms, six candidates were identified. These were subjected to vigorous tests to determine their nature; Lupus-TR-3 was quickly classified as a prime target.  A full analysis of the other candidates will form the subject of a separate paper. Four full transits and two half-transits (in egress phase) of Lupus-TR-3b were observed during our 53 nights. The photometry was used to determine the ephemeris, best-fitting model, and system parameters for comparison with the spectroscopic results.

\begin{centering}
\plotone{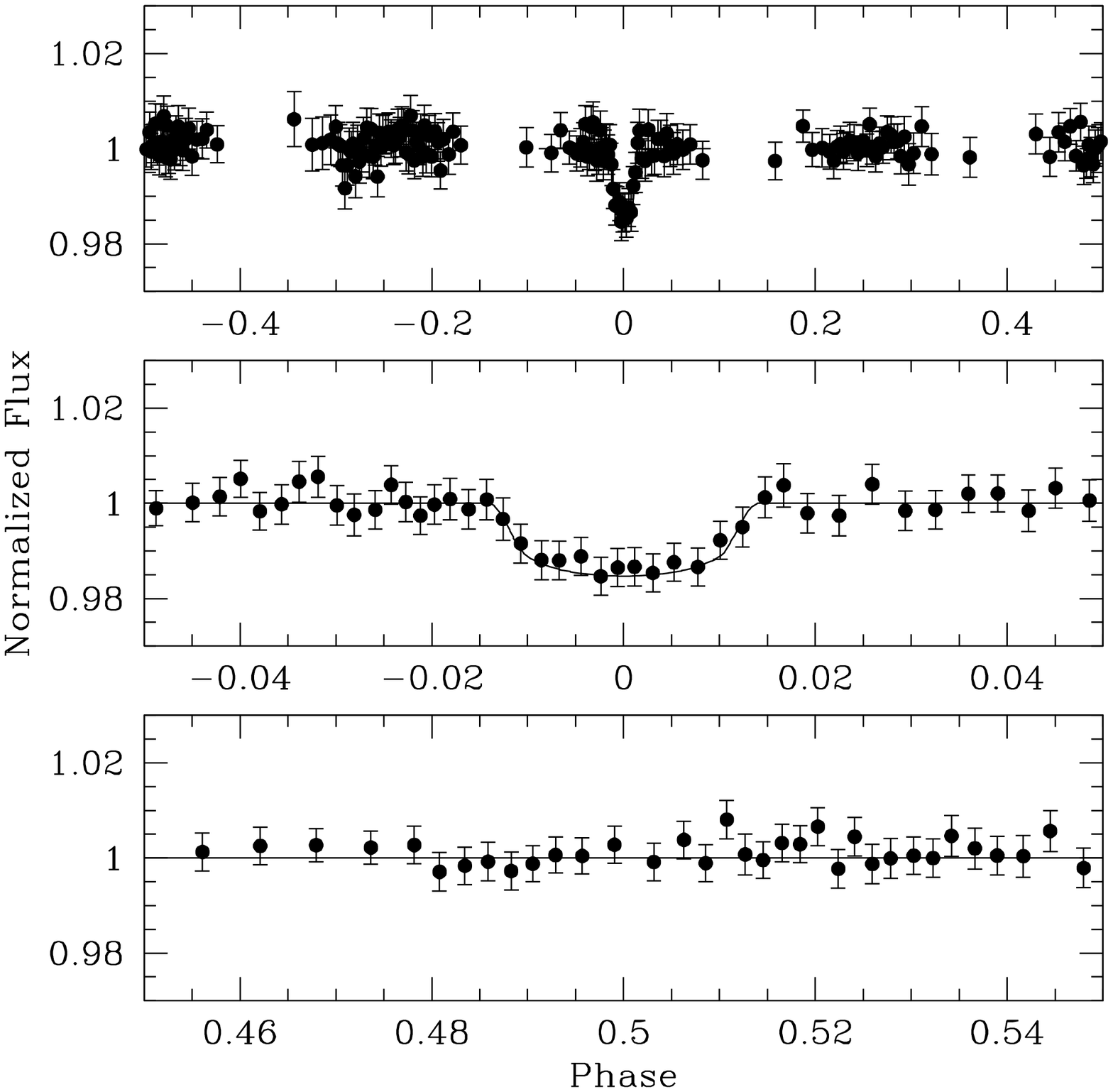}
\vskip -0.3cm
\figcaption{Differential, phase-wrapped Lupus-TR-3 photometry as a function of transit phase after application of {\sc SYSREM} \citep{T2005}. Data is binned using a 9-point binning, with the best-fitting model shown (middle and bottom) as a solid line. The bottom panel is centered on the time of anti-transit.\label{lc2}} 
\end{centering}
\vspace{0.15cm}

The transit photometry is shown in Fig.\space\ref{lc2}, covering the full range in phase and zoomed on the transit itself.  Also shown in Fig.\space\ref{lc2} is the best-fitting model assuming a circular orbit, a V+R quadratic limb-darkening law, and $R_{\ast}$=0.82 $\pm$ 0.05$R_{\odot}$ (see next section).  All system parameters derived from the photometric data and the best-fit model are given in Table\space1. The transit has observed duration $0.109 ^{+0.003}_{-0.007}$ days and maximum $\delta$F/F$=$0.013 $\pm$ 0.001. The {$\it{rms}$} residuals from the model are 7.48$\times$10$^{-3}$ mag. 

\vspace{-0.15cm}
\begin{center}
\footnotesize
\begin{tabular}{ll} \hline
$\rm{Parameter}$ & $\rm{Value}$ \\ \hline \hline 
Period & 3.91405 $\pm$ 0.00004 days\\ 
a & 0.0464 $\pm$ 0.0007 AU\\ 
T$_{\rm{center}}$ & 2453887.0818 $\pm$ 0.0013 HJD\\ 
R$_{\rm{p}}$ & 0.89 $\pm$ 0.07 R$_{\rm{J}}$\\ 
$i$ & $88.3 ^{+1.3}_{-0.8}$ deg\\ 
$K$  & 114 $\pm$ 25 m/s \\
M$_{\rm{P}}$ & 0.81 $\pm$ 0.18 M$_{\rm{J}}$ \\ 
$\rho$ & 1.4 $\pm$ 0.4 g/cm$^{3}$ \\ 
$\rm{e}$ & 0 (fixed)\\ \hline
\end{tabular}
\end{center}
\vspace{-0.1cm}
\footnotesize{Table 1: Lupus-TR-3b parameters.}
\normalsize

\vspace{-0.2cm}
\section{Characterizing Lupus-TR-3}
V and I magnitudes were obtained for all stars in the survey field using the ANU 1m telescope. In addition, spectroscopy was obtained with the DBS on the ANU 2.3m telescope to determine the stellar type (R$=$200,  per-pixel S/N$=$13), reduced using the "onedspec" package of IRAF. Wavelength calibration was performed using CuHe arcs taken throughout the same night. Five FEROS spectra on the MPG/ESO 2.2m, taken over four nights, ruled out a stellar binary at a few km/s level. The DBS spectrum was corrected for E(B$-$V)$=$0.182 mag reddening \citep{Sch1998} and compared to MILES \citep{SB2004} empirical spectra. The results indicate that the target is a K1V star with effective temperature 5000 $\pm$ 150K and log(g)$=$4.5 $\pm$ 0.5. With these constraints and the measured stellar V$-$I, \citet{Y2003} isochrones indicate a mass of 0.87 $\pm$ 0.04M$_{\odot}$ and radius of 0.82 $\pm$ 0.05R$_{\odot}$, assuming solar metallicity.  As a check, we examined \citet{Bert1994} isochrones, which produced a consistent mass of 0.83 $\pm$ 0.03M$_{\odot}$.  The Yi isochrone stellar mass and radius were used to derive R$_{\rm{p}}$ and M$_{\rm{p}}$. From the assumed isochrone luminosity, the distance is estimated at $\sim$2 Kpc. The stellar parameters are summarized in Table\space2.

\begin{center}
\footnotesize
\begin{tabular}{ll} \hline
$\rm{Parameter}$ & $\rm{Value}$\\ \hline \hline
R.A. (J2000.0) & 15$^{\rm{h}}$30$^{\rm{m}}$18.7$^{\rm{s}}$\\
DEC. (J2000.0) & $-$42$^{\circ}$58$'$41.5$"$\\ 
V & 17.4 $\pm$ 0.2\\ 
V-I & 0.859 $\pm$ 0.008\\ 
Spectral Type & K1V\\ 
T$_{\rm{eff}}$ & 5000 $\pm$ 150K\\
Mass & 0.87 $\pm$ 0.04M$_{\odot}$\\ 
Radius & 0.82 $\pm$ 0.05R$_{\odot}$\\ 
Dist & $\sim$2 Kpc\\  \hline 
\end{tabular}
\end{center}
\footnotesize{Table 2: Stellar parameters for Lupus-TR-3.}
\normalsize

\section{Radial Velocity Measurements}
The photometry and RV measurements for Lupus-TR-3b are available upon request. We obtained high-resolution spectroscopic observations using the MIKE spectrograph on the Magellan II (Clay) telescope at Las Campanas Observatory in June 2007.  MIKE had successfully confirmed the transiting planet orbiting OGLE-TR-113b \citep{K2004}, a somewhat brighter star (I=14.4) of a similar spectral type to Lupus-TR-3.  Using MIKE in its standard configuration and a 0.35$''$ slit, fifteen usable spectra were obtained over four nights, with resolutions of R$=$47,000 to 48,000, signal-to-noise of 5-10 per pixel, and wavelength ranges of 5000-9500\AA. Exposure times varied between 1400-2400s.  In addition, spectra were taken of three bright radial velocity standards selected to match the spectral type and class of Lupus-TR-3.  The target and one of the standards were observed at airmass near unity each night.  The seeing during the run varied from 0.7$''$ to 1.05$''$ and there is no correlation between radial velocity and seeing, minimizing the chance of any variable contaminating spectra.

These data were reduced using standard IRAF\footnote{IRAF is distributed by the National Optical Astronomy Observatories, which are operated by the Association of Universities for Research in Astronomy, Inc., under cooperative agreement with the National Science Foundation.}  tasks and the {\it mtools\/} package (for correcting the tilted slits in MIKE spectra). ThAr arcs, taken before and after each exposure, were used to wavelength calibrate the spectra.  Radial velocities (RV) for each echelle order were computed by cross-correlating the Lupus-TR-3 spectra with one long exposure of K5V HD~219509 that consistently produced the most stable, low-scatter results for the target and HD~126053, another standard observed at approximately the same times and airmass as Lupus-TR-3.

The final results, shown in Fig.\space\ref{mikeRV}, use 18 orders selected for high signal-to-noise ($S/N$) and low telluric contamination; the scatter in these orders was used as an estimate of the uncertainity in the radial velocity.  Only the period of the transit was fixed in fitting the radial velocity measurements. Assuming a circular orbit, the best fit sinusoidal variation, detected at 2.4$\sigma$ compared to the null hypothesis, has an amplitude of K$=$114 $\pm$ 25~m/s and a phase consistent with the transit ephemeris, to within 0.13d, well within fit uncertainties. Uncertainties were calculated from the covariance matrix of the fitted parameters. 

\begin{centering}
\plotone{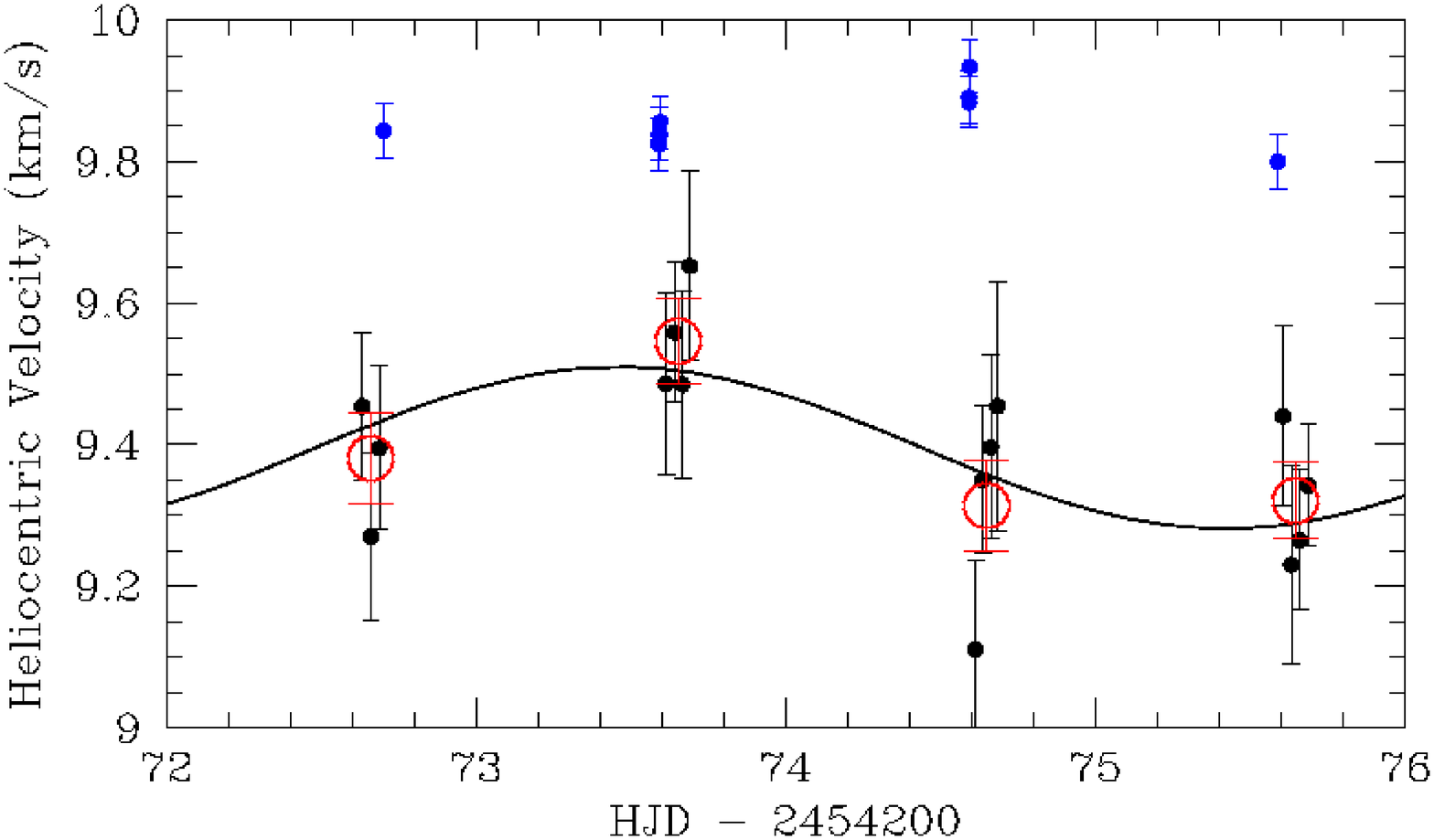} 
\figcaption{RV for Lupus-TR-3 (lower points) and the bright standard HD~126053 (upper, offset points) as a function of Heliocentric Julian Date.  Error bars are determined from the actual scatter in the echelle orders used. Larger open circles are uncertainty-weighted nightly averages.\label{mikeRV}}
\end{centering}
\vspace{0.3cm}

The RV signal is small and its detection significance rather low, which is not surprising given the faintness of Lupus-TR-3.  Detecting the signal required an exceptional wavelength solution with an ${\it{rms}}$ of 0.0011 pixels, rejection of echelle orders susceptible to terrestrial contaminants, observations at nearly constant and very low airmass, and high-signal-to-noise RV standards of similar spectral type.  We note that the scatter in the points taken on any given night can be taken as a proxy for the reliability of our measurements.  This scatter over 2-3 hour time frames is consistent with the uncertainties displayed on individual points in Fig.\space\ref{mikeRV}, which were, in turn, calculated from the actual scatter across individual echelle orders for that spectrum. The resulting planetary mass is 0.81 $\pm$ 0.18 M$_{\rm{J}}$, assuming a 0.87 $\pm$ 0.04M$_{\odot}$ host star.  Fixing the phase to match the transit data produced a nearly identical planetary mass (of even greater significance), as did different choices for the echelle orders used in the analysis.

\section{Ruling out Blended Binaries}
\subsection{Image Deconvolution with DECPHOT}
In principle, a transit signal can be caused by a faint eclipsing binary system that is located close enough to the target to be blended within the target's point spread function (PSF), producing the observed transit and usually no detectable radial velocity periodicity.  On brighter targets, line bisector analysis can be used to determine if a blend is present. Our data show no sign of in-phase bisector variations, but given the low signal-to-noise of the spectra, the constraint on blended binaries is inconclusive.

We therefore also tested for blends by performing a DECPHOT analysis \citep{Mag2007,G2007a} over one night in order to effectively ``de-blend'' our ANU 1m images of Lupus-TR-3. Two faint companions (B and C) were detected inside the ~2$''$ PSF of the target on the ANU 1m, one 4.1 mag fainter than Lupus-TR-3 located ~1.7$''$ away (B), and the other 5.6 mag fainter at 2.2$''$ distance (C). Star C is too faint to be a binary masquerading as a planet. 

\vspace{-0.15cm}
\subsection{Y-Band Imaging}
We also obtained high-spatial resolution images in 0.68$''$ seeing taken in the Y-band (1.035 microns) with PANIC (0$''$.125 pixels) on Magellan I, to further rule out the possiblity of a blend and to check the DECPHOT deconvolution. We obtained 18 frames with exposure times of 120s each, and the best quality frames were combined to make a deep image of the field (Fig.\space\ref{nir}).  

\begin{centering}
\plotone{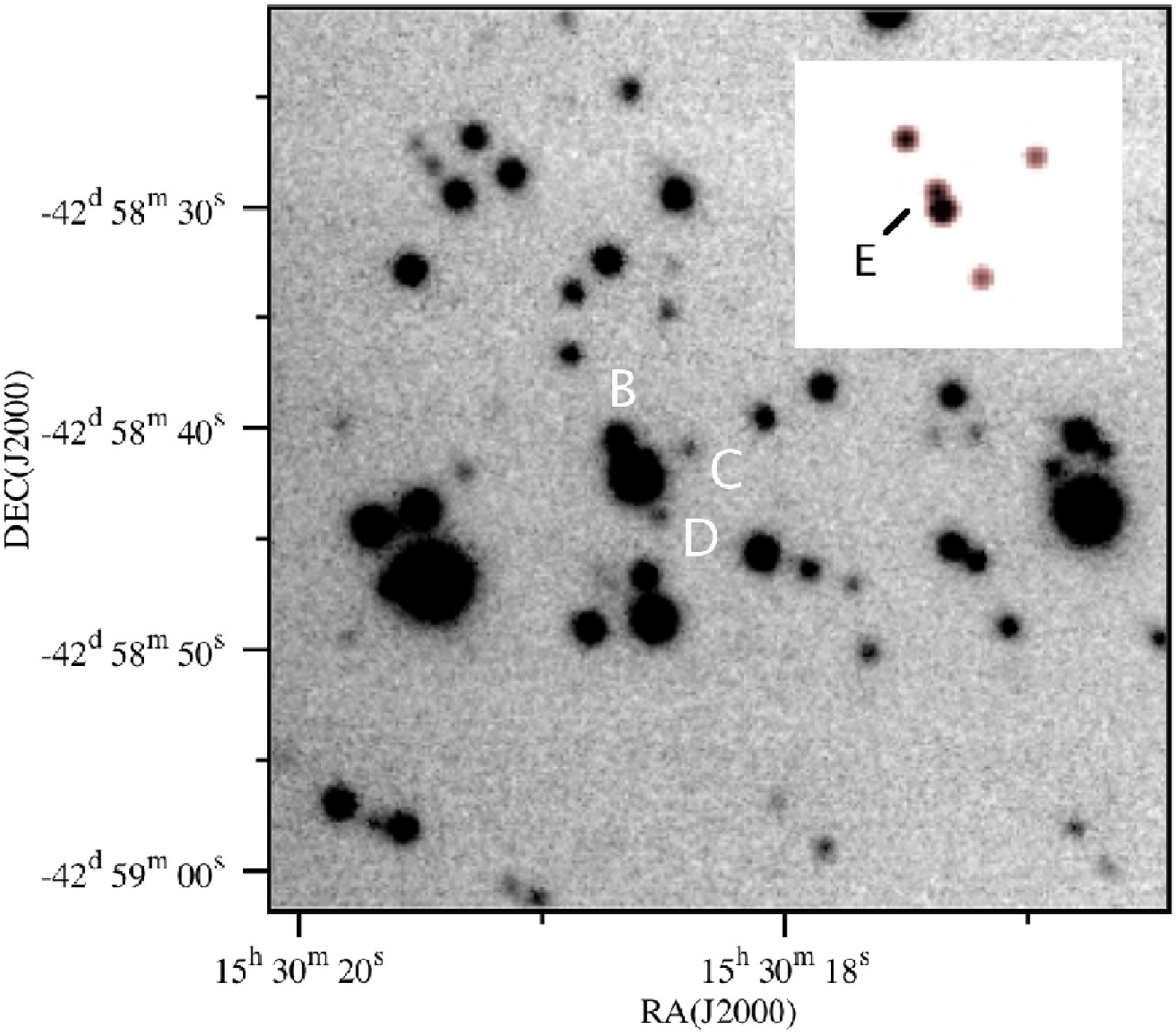}
\figcaption{Y-band 40$\times$45 arcsec$^2$ PANIC imagery for Lupus-TR-3 (left of center). Three blended neighbors are revealed, the two detected with DECPHOT on our ANU 1m images (B and C) and a third (D). Inset is the post-DECPHOT result for the target, revealing a fourth blended companion, star E.
\label{nir}}
\end{centering}
\vspace{0.2cm}

This image confirms the presence of the two neighbors found using DECPHOT on our 1m images at the same separations, with Y-band magnitude differences of 2.5 for star B and 4.2 for star C. In addition, one other close neighbor (D) was detected 4.3 magnitudes fainter in the Y-band than Lupus-TR-3 at a separation of 1.8$''$. Star D is too faint in V+R to harbor a feasible confusing eclipse signal, and in any event we have already shown that the transit is strongly associated with the position of the target. 

We also performed a DECPHOT analysis of the PANIC image, seen superimposed on Fig.\space\ref{nir}, which revealed another star, star E, 0.4$''$ away from the target. This could constitute a blended binary. Star E is not seen in our deconvolved V+R images, indicating a V magnitude fainter than 21, contributing $\le$0.036 magnitudes to the target brightness. It is 2.8 magnitudes fainter than Lupus-TR-3 in the Y-band, indicating a red spectral type. If star E is a binary that mimics the planetary hypothesis, it must harbor eclipses with a depth of at least 36$\%$. Planned observations will test this possibility.

DECPHOT was then used to analyze 73 images from one night containing a well-sampled transit. Light curves were created for the target and star B, to determine with which star, and in what location, the transit signal was produced. The result is shown in Fig.\space\ref{lc1}. The transit signal clearly occurs at the target location and with the same depth and duration. This rules out star B as a binary, with C and D being ruled out by their angular distance and faintness. Only star E remains as a blend candidate.

\begin{centering}
\plotone{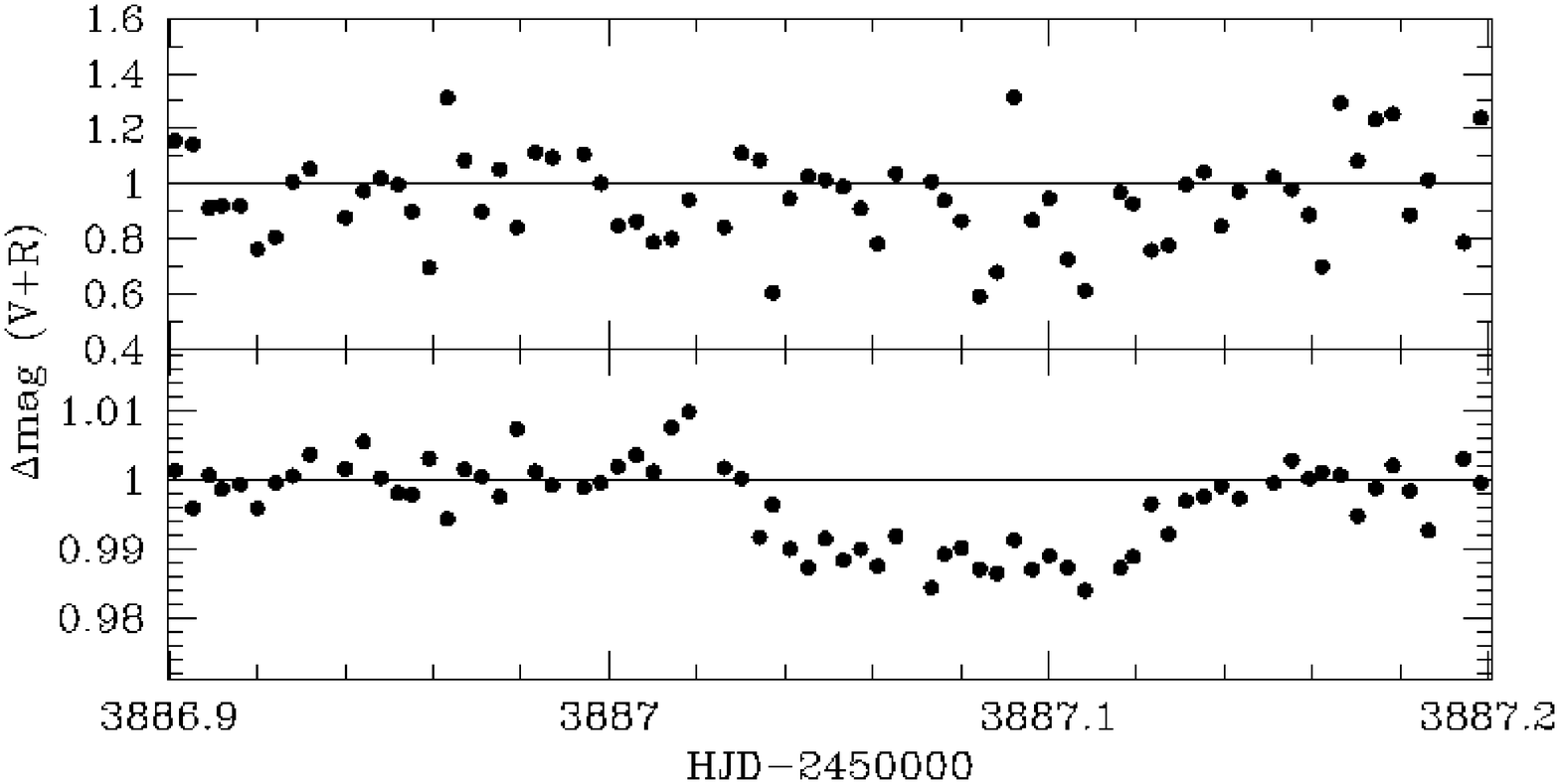}
\figcaption{Post-DECPHOT in-transit photometry for Lupus-TR-3 (bottom) and star B (top). The signal is strongly associated with the the target location, ruling out star B as a binary. Star C is ruled out due to its larger angular distance, and D is too faint to harbor any eclipse. 
\label{lc1}}
\end{centering}

\subsection{Modeling Possible Binary Configurations}
A large number of stellar binaries were modeled to examine if star E could be a blended binary. The tested binary models spanned 21 spectral types between A7 and M3 eclipsing at 41 different impact parameters. Our well-determined transit morphology places strict constraints on the nature of any prospective blend, as Lupus-TR-3 displays a large flat-bottomed transit, which depends most significantly on the ratio of radii between the components and the impact parameter of the transit and less importantly on the orbital eccentricity and total system mass.

The best fit blend scenario (A7 primary and M3 secondary with a 0.134 mag central eclipse, hence 4.7 mags fainter than the target) can be ruled out based on the transit duration. An eccentricity of $\geqslant$0.56 would be required to reduce the circular orbit duration of 0.21d to the observed 0.1d event. Non-zero eccentricities are not observed in any binary systems with such short periods \citep{abt2006}. In any case the redness and V+R faintness of star E makes it unlikely to contain an A-type primary.

Binary systems with larger primaries can also mimic the observations, albeit less ably, and can also be ruled out based on the more extreme eccentricities required for larger, earlier spectral types primaries. Also, hotter and larger stars increase the probability that ellipsoidal variations will be observed in our data. A blended star with a rare small companion close to the hydrogen burning limit (ie: OGLE-TR-122b, \citealt{Pont2005}) can also be rejected as it would have to be of very comparable brightness to Lupus-TR-3 to produce the observed transit depth (such a system would have similar undiluted transit depth). Such a star is not observed in our DECPHOT analysis.

In summary, in the unlikely event that the eclipse in Lupus-TR-3 is caused by a blend, only star E can be responsible. However its redness and faintness argue against the A+M binary configurations most capable of reproducing our observations. Our data cannot rule out the distant possibility that star E is an usual binary system or a stellar companion to Lupus-TR-3 itself. The other three close neighbors are ruled out using DECPHOT. 

\section{Conclusions}
We announce the detection of Lupus-TR-3b, a strong candidate for a transiting Hot Jupiter orbiting a V$=$17.4 K1V star with a period of 3.91405d. Our observations are consistent with a planet of radius R$_{\rm{p}}$ $=$ 0.89 $\pm$ 0.07 R$_{\rm{J}}$ (assuming a Jovian equatorial radius of 71,492 kms). MIKE radial velocities reveal a 2.4$\sigma$ low-amplitude variation in phase with the transit ephemeris, yielding a planetary mass of 0.81 $\pm$ 0.18M$_{\rm{J}}$ and thus density of 1.4 $\pm$ 0.4g/cm$^{3}$, equivalent to that of Jupiter. 

Deep imaging, image deconvolution techniques and extensive binary modeling argue strongly against a blended or triple configuration for the system that can reproduce the photometric observations. With current data, we cannot rule out the unlikely possibility that the nearby faint, red star E is a blended binary responsible for the transit, but this is testable with minimal amounts of planned high resolution imaging. The low-mass, Hot Jupiter Lupus-TR-3b is an example of the extraordinary discoveries that await if innovative computational methods and time-efficient follow-up observations are brought to bear on candidates selected from wide and deep transit surveys.

\acknowledgements The authors wish to thank Grant Kennedy and Karen Lewis for assistance during the original observing runs; Ken Freeman for the ANU 2.3m spectrum; Andre M\"uller for the FEROS spectra; Nidia Morrell, Miguel Roth, and Chris Burns for the Magellan Y-band images of Lupus-TR-3, and the anonymous referee. We acknowledge financial support from the Access to Major Research Facilities Programme which is a component of the International Science Linkages Programme established under the Australian Government's innovation statement, Backing Australia's Ability.

\end{document}